\documentclass[aps,pre,twocolumn,superscriptaddress]{revtex4}

\usepackage{amsfonts,amssymb,amsmath,latexsym,epsfig,wasysym} 
\usepackage[sort&compress]{natbib}

\usepackage{color}
\definecolor{bluecolor}{rgb}{0,0.,1.}

\definecolor{redcolor}{rgb}{.7,0.,0.}

\usepackage{framed}

{\begin{snugshade}\begin{quote}}
{\hfill\end{quote}\end{snugshade}}

\definecolor{shadecolor}{rgb}{0.9,0.9,0.9}

\begin{document}

\title{Statistical laws in linguistics}

\author{Eduardo G. Altmann} 
\affiliation{Max Planck Institute for the Physics of Complex Systems, 01187 Dresden, Germany}
\author{Martin Gerlach} 
\affiliation{Max Planck Institute for the Physics of Complex Systems, 01187 Dresden, Germany}

\begin{abstract}
Zipf's law is just one out of many universal laws proposed to describe statistical regularities in language.
Here we review and critically discuss how these laws can be statistically interpreted, fitted, and tested (falsified).
%
%
The modern availability of large databases of written text allows for tests with an
unprecedent statistical  accuracy and also a characterization of the fluctuations
around the typical behavior.
We find that fluctuations are usually much larger than expected based on
simplifying statistical assumptions (e.g., independence and lack of correlations between
observations).  
These simplifications appear also in usual statistical tests so that the large
fluctuations can be erroneously interpreted as a falsification of the law. Instead, here
we argue that  linguistic laws are only meaningful (falsifiable) if accompanied by a model
for which the fluctuations can be computed (e.g., a generative model of the text). 
The large fluctuations we report show that the constraints imposed by linguistic laws on
the creativity  process of text generation are not as tight as one could expect.

\noindent Proceedings of the {\it Flow Machines Workshop: Creativity and
  Universality in Language}, Paris, June 18 to 20, 2014.
\end{abstract} 

\maketitle

\tableofcontents



\section{Introduction}\label{sec.intro}

\begin{quote}
``{\it ...'language in use' cannot be studied without statistics''} Gustav
Herdan (1964)~\cite{Herdan}
\end{quote}

In the past 100 years regularities in the frequency of text constituents have
been summarized in the form of {\it linguistic laws}.
For instance,  Zipf's law 
states that the frequency $f$ of the $r$-th most frequent word in a text is inversely
proportional to its rank: $f \propto
1/r$~\cite{Zipf}.  This and other less famous linguistic laws are one of the main objects
of study of {\it quantitative  linguistics}~\cite{QL,Series,Glottopedia,TrierWebpage,Baayen,Zanette}.

Linguistic laws have both theoretical and practical importance.  They provide insights
on the mechanisms of text (language, thought) production and are also crucial in applications of
statistical natural language processing (e.g., information retrieval). 
Both the generative and data-analysis views of linguistic laws are increasingly important
in modern applications. Data-mining algorithms profit
from accurate estimations of the vocabulary size of a collection of texts (corpus), e.g.,
through Heaps' law discussed in the next section. Methods for the automatic  generation of
natural language can profit from knowing the linguistic laws underlying usual texts. For
instance,  linguistic laws may be included as (additional) constraints in the
space of  possible (Markov generated) texts~\cite{Barbieri} and can thus be considered as  
constrains to the creativity of authors.

\begin{figure*}[!bt]
\centering
\includegraphics[width=1.8\columnwidth]{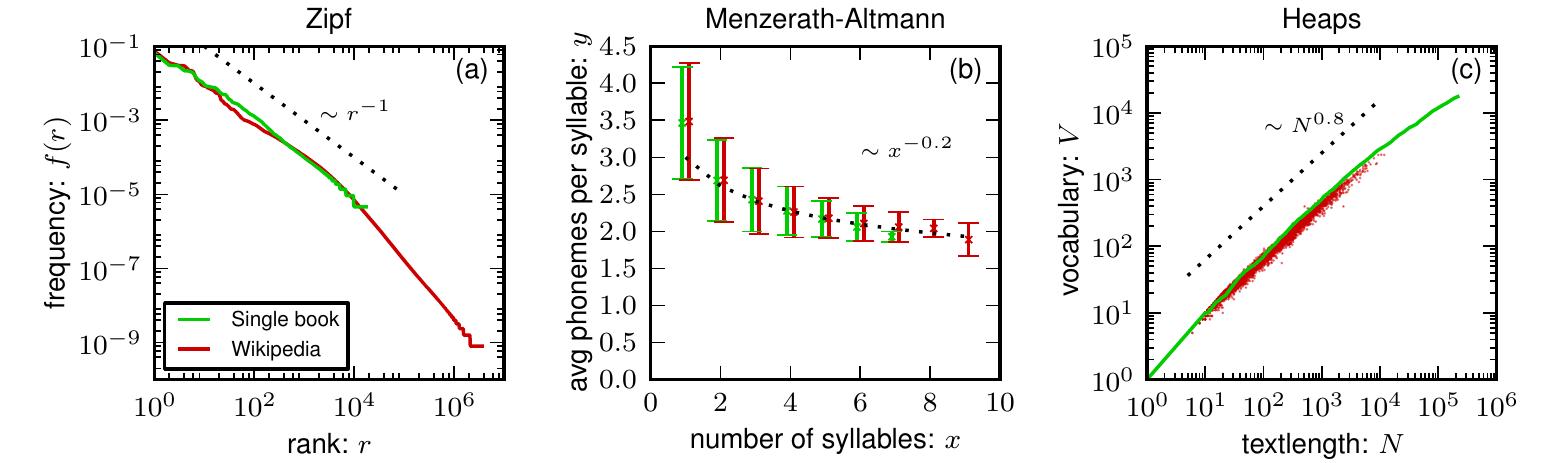}
\caption{Examples of linguistic laws: (a) Zipf, (b) Menzerath-Altmann, and
  (c) Heaps laws.  Data from one book (green, Moby Dick by H. Melville) and for the
  English Wikipedia (red) are shown. Dotted (black) lines
  are the linguistic laws with arbitrary parameter, chosen for visual
  comparison (see Appendix for details).}
\label{fig.1}
\end{figure*}

Besides giving an overview on various examples of linguistic laws
(Sec.~\ref{sec.observations}), in this paper we focus on their probabilistic 
interpretation (Sec.~\ref{sec.interpretation}), we discuss different statistical methods
of data analysis (Sec.~\ref{sec.data}), and the possibilities of connecting different laws
(Sec.~\ref{sec.relation}). The modern availability of large text databases allows for an
improved view on linguistic laws that requires a careful discussion of their
interpretation. Typically, more data confirms the observations motivating 
the laws -- mostly based on visual inspection -- but makes increasingly difficult for the
laws to pass statistical tests designed to evaluate their validity. This leads to a seemingly contradictory situation:
while the laws allow for an estimation of the general behavior 
(e.g., they are much better than alternative descriptions), they are strictly-speaking 
falsified.  The aim of this contribution is to present this problem and discuss alternative
interpretations of the results. We argue that the statistical analysis of linguistic laws
often shows long-range correlations and large (topical) fluctuations. We conclude that null
models accounting for these observations are often ignored yet crucial in the tests of
the validity of linguistic laws.

\section{Examples and Observations}\label{sec.observations}

An insightful introduction to Linguistic Laws is given in Ref.~\cite{Glottopedia} by
 K\"ohler, who distinguishes between three kinds of laws as follows:

\begin{enumerate}
\item ``The first kind takes the form of probability distributions, i.e. it makes predictions
about the number of units of a given property.''
\item ``The second kind of law is called the functional type, because these laws link two (or more) variables, i.e. properties.''
\item ``The third kind of law is the developmental one. Here, a property is related to
  time.'' (time may be measured in terms of text length)
\end{enumerate}
 We use the term linguistic law to denote quantitative relationships between
measurements obtained in a written text or corpus, in contrast to syntactic rules and to
phonetic and language-change laws (e.g., Grimm's law). 
We assume that the laws make statements about individual texts (corpus) and are
 exact in an appropriate 
limit (e.g., large corpus)\footnote{While some of the laws clearly intend to speak about the
language as whole, in practice they are tested and motivated by observations in specific
texts which are thus implicitly or explicitly assumed to reflect the language as a
whole. }.
 Each law contains parameters which we denote by Greek letters $\alpha,
 \beta, \gamma$, and  often refer to the frequency $f(q)$ of a quantity $q$ in the text
 (with $\sum_q f(q)=1$). Probabilities are denoted by $P(q)$.

Next we discuss in detail one representative example of each of the three types of laws
mentioned above: Zipf,
Menzerath-Altmann, and Heaps laws, respectively, see Fig.~\ref{fig.1}.

\begin{table*}
  \caption{List of linguistic laws.}\label{tab.laws}
  \centering
  \begin{tabular}{|l|c|c| c|}
\hline
    {Name of the law}
&{Observables}
&{Functional form}
&{References}\\
\hline
{Zipf}
&{$f$: freq. of word $w$; $r$: rank of $w$ in $f$}
&{$ f(r)=\beta_Z r^{-\alpha_Z}$}
&{\cite{Zipf,Simon,Mitzenmacher2004,Newman,Li,Montemurro,Piantadosi2014}}\\
{Menzerath-Altmann}
&{$x:$ length of the whole; $y:$ size of the parts}
&{$y= \alpha_M x^{\beta_M} e^{-\gamma_M x}$}
&{\cite{Altmann1980,Cramer2005}}\\
{Heaps}
&{$V:$ number of words; $N:$ database size}
&{$V\sim N^{\alpha_H}$}
&{\cite{Herdan,Egghe,Lu,Petersen,PRX,Francesc,NJP}}\\
\hline
{Recurrence}
&{$\tau:$ distance between words }
&{$P(\tau)\sim \exp{(\alpha \tau)}^\beta$}
&{\cite{Zipf,Altmann2009,Lijffijt2011}}\\
{Long-range correlation}
&{$C(\tau)$: autocorrelation at lag $\tau$}
&{$C(\tau) \sim \tau^{-\alpha}$}
&{\cite{Mandelbrot73,Schenkel93,PNAS}}\\
{Entropy Scaling}
&{$H:$ Entropy of text with blocks of size $n$}
&{$H \sim \alpha n^\beta + \gamma n$}
&{\cite{Ebeling1994,Debowski2006}}\\
{Information content}
&{$I(l):$ Information of word with length $l$}
&{$I(l) = \alpha + \beta l$}
&{\cite{Zipf,Piantadosi2011}}\\ 
{Taylor's law}
&{$\sigma$: standard deviation around the mean $\mu$}
&{$\sigma \sim \mu^\alpha$}
&{\cite{NJP}}\\
{Networks}
&{Topology of lexical/semantic networks}
&{various}
&{\cite{Sole10,Choudhurry09,Baronchelli13,Cong14}}\\

\hline
  \end{tabular}
\end{table*}

\subsection{Zipf's law}

Zipf's law is the best known linguistic law
(see, e.g., Ref.~\cite{Mitzenmacher2004} for historical references). In an early and simple formulation,  
it states that if words (types) are ranked according to their frequency of appearance $r=1,2,
\ldots, V$, the frequency $f(r)$ of the r-th word (type) scales with the rank as
\begin{equation}\label{eq.zipf}
f(r)=\frac{f(1)}{r},
\end{equation}
where $f(1)$ is the frequency of the most frequent word. The above expression cannot hold
for large $r$ because for any $f(1)>0$, there is an $r^*$ such that $\sum_{r=1}^{r^*} f(1)/r
> 1$. Taking also into account that $f(1)$ may not be the best proportionality factor,
a modern version of Zipf's law is
\begin{equation}\label{eq.zipf2}
f(r)=\frac{\beta_Z}{r^{\alpha_Z}},
\end{equation}
with $\alpha_Z \ge 1$, see Fig.~\ref{fig.1}(a). The analogy with other processes showing
fat-tailed distribution motivates the alternative formulation
\begin{equation}\label{eq.zipf3}
P(f)=\frac{\beta^\dagger_Z}{f^{\alpha^\dagger_Z}},
\end{equation}
where $P(f)$ is the fraction of the total number of words (probability) that have
frequency $f$. Formulations~(\ref{eq.zipf2}) and~(\ref{eq.zipf3}) can be mapped to each
other with $\alpha^\dagger=1+1/\alpha$~\cite{Mandelbrot61,Newman,Mitzenmacher2004}.

\subsection{Menzerath-Altmann law}
The Menzerath-Altmann law received considerable
attention after the works of
Gabriel Altmann~\cite{Series,Glottopedia,TrierWebpage,Altmann1980}.  
Menzerath's  general (qualitative) statement originating from his observations about phonemes is that ``the
greater the whole the smaller its parts''. The quantitative law intended to describe this
observation is~\cite{Altmann1980}
\begin{equation}\label{eq.menzerath}
y= \alpha_M x^{\beta_M} e^{-\gamma_M x},
\end{equation}
where $x$ measures the length of the whole and $y$ the (average) size of the parts. One
example~\cite{Altmann1980} is obtained computing for
each word $w$ the number of  syllables $x_w$ and the number of phonemes $z_w$.  The length of the word (the whole) is measured by the number of
syllables~$x_w$, while the length of the parts is measured for each word as the average number
of phonemes per syllable $y_w=z_w/x_w$.  The comparison to the law is made by averaging
$y_w$ over all words $w$ with $x_w=x$, see Fig.~\ref{fig.1}(b).  The ideas of
Menzerath-Altmann law and Eq.~(\ref{eq.menzerath}) have been extended and applied to a
variety of problems, see Ref.~\cite{Cramer2005} and references therein.

\subsection{Heaps' law}
Heaps' law states that the number of different words $V$ (i.e., word types) scales with
database size $N$ measured in the total number of words (i.e., word tokens) as~\cite{Herdan,Egghe} 
\begin{equation}\label{eq.heaps}
V\sim N^{\alpha_H}.
\end{equation}
In Fig.~\ref{fig.1}(c) this relationship is shown in two different representations. For a
single book, the value of $N$ is increased from the first word (token) until the end of
the book so that $V(N)$ draws a curve. For the English Wikipedia, each article is
considered as a separate document for which $V$ and $N$ are computed and shown as dots.

\bigskip 

The non-trivial regularities and the similarity between the two disparate databases found
for the three cases analyzed in Fig.~\ref{fig.1} strongly suggest that the three
linguistic laws summarized above capture important
properties of the structure of texts. 
Additional examples of linguistic laws are
listed in Tab.~\ref{tab.laws}, see also the vast literature in quantitative
linguistics~\cite{QL,Series,TrierWebpage,Glottopedia}. 
The (qualitative) observations reported above motivate us to search for quantitative
analysis that match the requirements of applications and the accuracy made possible
through the use of large corpora. 
The natural questions that
we would like to address here are: Are these laws true (compatible with the observations)? How to determine their  parameters? How
much fluctuations around them should be expected (allowed)? Are these laws related to each
other? Before addressing these questions we discuss {\it  how should one interpret linguistic laws}.

\section{Interpretation of Linguistic Laws}\label{sec.interpretation}

In Chap.~26 {\it Text Laws} of Ref.~\cite{QL},
H\v{r}ebi\v{c}ek argues that 
\begin{quotation}
``...the notion {\it law} (in the narrower sense {\it scientific law}) in linguistics and
especially in quantitative linguistics  ... need not obtain some special comprehension
different from its validity in other sciences. Probably, the best delimitation of this
concept can be found in the works by the philosopher of scientific knowledge Karl Raimund
Popper...'' 
\end{quotation}
This view is also emphasized by K\"ohler in Ref.~\cite{Glottopedia}, who distinguishes {\it laws} from {\it rules} and states that a {\it `` significant difference is that rules can be violated - laws (in the scientific sense) cannot.''}. 

Such a straight-forward identification between linguistic and scientific laws masks the
central role played by statistics (and probability theory) in the interpretation of linguistic laws. To see this,
first notice that these laws do not directly affect the
production of (grammatically and semantically) meaningful sentences because they typically
involve scales much larger or shorter than a sentence.  It is thus not difficult to be
convinced  that a creative and persistent daemon~\footnote{A relative of Maxwell's
   Daemon known from Thermodynamics.}, trained in the techniques of {\it constrained
   writing}~\cite{WikipediaConstrainedWriting}, can generate understandable and arbitrary long
 texts which deliberately violate any single law mentioned above.  In a strict
 Popperian sense, a single of such demonic texts would be sufficient to falsify the proposed
 laws. Linguistic  laws  are thus different from syntactic rules and require a different
 interpretation than,  e.g.,  the laws of classical  Physics. 

The central role of statistics in Quantitative Linguistics was emphasized by its founding
father Gustav Herdan:
\begin{quotation}
``The distinction between language laws in the conventional sense and statistical laws of
language corresponds closely to that between the classical laws of nature, or the physical
universe, and the statistical laws of modern physics.''~\cite{Herdan}
\end{quotation}
Altmann, when discussing Menzerath law~\cite{Altmann1980}, also emphasizes that ``this law
is a stochastic one'', and K\"ohler~\cite{QL} refers to the concept of stochastic hypothesis.
There are at least two instances in which a statistical interpretation should be included:
\begin{enumerate}
\item  In the statement of the law, e.g., in Zipf's law the probability of  finding a word
  with    frequency $f$ decays as $P(f) \sim f^{-\alpha^\dagger_z}$.  
\item In the interpretation of the law as being {\it typical} in a collection of
  texts, e.g.,  in Heaps' law the vocabulary $V$ of a (typical) text of size $N$ is $V\sim
  N^\alpha_H$. 
\end{enumerate}
The demonic texts mentioned above would be considered {\it untypical} (or highly unlikely). 
Statistical laws in at least one of these senses are characteristic not only of modern Physics,
as pointed out by Herdan,
but also of different areas of natural
and social sciences: Benford's law predicts the frequency of the first digit of
numbers appearing in a corpus and the
Gutenberg-Richter law determines the frequency of earthquakes of a given magnitude.
The analysis of these laws, including possible refutations, have to be done through
rigorous statistical methods, the subject of the next section.
Important aspects of linguistic laws not discussed in detail in this Chapter include: (i)
the universality and variability of parameters of  linguistic laws (e.g., across different
languages~\cite{Petersen,PRX,Amancio,Cong14,Ger}, as a function of
size~\cite{Bernhardsson2010} and degree of mixture of the corpus~\cite{Dodds}, styles~\cite{Schenkel93}, and age of speakers~\cite{Baixeries}); and
(ii) the relevance and origins of the laws. This second point was intensively debated
for Zipf's law~\cite{Zanette,Jaeger,Piantadosi2014}, with quantitative 
approaches based on stochastic processes 
-- e.g., the Monkey typewriter model~\cite{Newman}), rich-get-richer
mechanisms~\cite{Simon,Mitzenmacher2004,Montemurro,Newman,PRX} -- and on optimization principles -- e.g., between
speaker and  hearer~\cite{Zipf,Corominas-Murtra,Ferrer-i-Cancho14} or entropy maximization~\cite{Marsili13,Peterson13}.

\section{Statistical Analysis}\label{sec.data}

In Sec~\ref{sec.observations} we argued in favor of linguistic laws by showing a graphical
representation of the data (Fig.~\ref{fig.1}). The widespread availability of large
databases and the applications of linguistic laws require and allow for
a more rigorous statistical analysis of the results. To this end we assume the linguistic
law can be translated in a precise mathematical statement about a curve or distribution. This distribution has a set of parameters and observations.
Legitimate questions to be addressed are:

\begin{enumerate}
\item[(1)] {\bf Fitting.} What are the best parameters of the law to describe a given data? 
\item[(2)] {\bf Model Comparison.} Is the law better than an alternative one? 
\item[(3)] {\bf Validity.} Is the law compatible with the observations? 
\end{enumerate}
These points are representative of statistical analysis performed more generally and
should preceed any more fundamental discussion on the origin and importance of a specific
law. Below we discuss in more details how each of the three points listed above has been
and can be addressed in the case of linguistic laws.

\subsection{Graphical approaches}\label{ssec.graphical}

Visual inspection and graphical approaches were the first type of analysis of linguistic
laws and are still widely used. One simple and still very popular fitting
approach is least squares (minimize the squared distance between data and models). Often this 
is done in combination with a transformation
of variables that maps the law into a straight line (e.g., using logarithmic scales in the
axis or taking the logarithm of the independent and dependent variable in the Zipf's
and Heaps' laws). These transformations are important to visually detect patterns and are valuable part of any data analysis. However,
they are not appropriate for a quantitative analysis of the data. The problem of fitting
straight lines in log-log scale
is that least-square fitting assumes an uncertainty (fluctuation) on each point that is
independent, Gaussian distributed, and equal in size for all fitted points. These
assumptions are usually not justified (see, e.g., Refs.~\cite{Goldstein04,Bauke07} for the case of fitting power-law distributions),
while at the same time the uncertainties are modified through the transformation of variables (such as using the log scale). 
Furthermore, quantifying the goodness-of-fit by using the correlation-coefficient $R^2$ in these scales is insufficient to
evaluate the validity of a given law. A high quality of the fit indicates a high correlation between
data and model, but is unable to assign a probability for observations and thus it is not
suited for a rigorous test of the law.

\subsection{Likelihood methods}

A central quantity in the statistical analysis of data is the likelihood
$\mathcal{L}({\bf x};\vec{\alpha})$ that the data ${\bf x}$ was generated by the model
(with a set of parameters ${\bf \alpha}$).

\paragraph*{(1) Fitting}\label{ssec.fit}
When fitting a model (law) to data the approach is to tacitly assume its validity and then search
for the best parameters to account for the data. It corresponds to a search in the
(multidimensional) 
parameter space ${\bf \alpha}$ of the law for the value $\hat{\bf \alpha}$ that
maximize $\mathcal{L}$.

In laws of the first kind -- as listed in Sec.~\ref{sec.observations} -- the quantity to
be estimated from data is a probability distribution
$P({\bf x};{\bf \alpha})$. The probability of an observation $x_j$ is thus given
by 
$P(x_j;{\bf \alpha})$. Assuming that all $J$ observations are independent, the best
parameter estimates $\hat{\bf \alpha}$ are the values of ${\bf \alpha}$ that
maximize the log-likelihood
\begin{equation}\label{eq.likelihood}
\log_e \mathcal{L}= \log P(x_1, x_2, \ldots x_J;{\bf \alpha}) = \sum_{j=1}^J \log
\tilde{P}(x_j; {\bf \alpha}),
\end{equation}
The need for Maximum Likelihood (ML) methods when fitting power-law
distributions (such as Zipf's law) has been emphasized in many recent
publications.  We refer to the review article Ref. \cite{Clauset2009} and references
therein for more details, and to Ref.~\cite{Deluca13} for fitting truncated distributions
(e.g., due to cut-offs).

In laws of the second and third kind -- as listed in Sec.~\ref{sec.observations} -- the
quantity to be described $y$ is a function $y=y_g({\bf x};\vec{\alpha})$. Fitting requires
assumptions regarding the possible fluctuations in $y({\bf x})$.  One possibility is to assume
Gaussian fluctuations with a standard deviation $\sigma({\bf x})$. In this case, assuming
again that the observations ${\bf x}$ are independent~\cite{Hastie09}
\begin{equation}\label{eq.L2}
\log_e \mathcal{L} \sim -\sum_j \left(\frac{y({\bf x}_j)-y_g({\bf x}_j)}{\sigma({\bf x}_j)}\right)^2,
\end{equation}
where the sum is over all observations $j$. The best estimated parameters
$\hat{\vec{\alpha}}$ are obtained minimizing $\chi^2=\sum_j (\frac{y({\bf x}_j)-y_g({\bf x}_j)}{\sigma({\bf x}_j)})^2$, which
maximizes~(\ref{eq.L2}). Least-squares fitting is equivalent to Maximum-Likelihood fitting
only in the case of constant $\sigma$ (independent of ${\bf x}$)~\cite{Hastie09}.

\paragraph*{(2) Model Comparison}
The comparison between two different functional forms of the law ($m1$ and
$m2$) is done comparing their likelihoods, e.g.,  through the log-likelihood ratio $\log_e
\mathcal{L}_{m1}/\mathcal{L}_{m2}$~\cite{Burnham02}. A value $\log_e
\mathcal{L}_{m1}/\mathcal{L}_{m2}=1$  ($-1$) means it is
$e^1=2.718\ldots$ times more (less) likely that the data was generated by function $m1$
than by function $m2$. If the two models have a
different number of parameters, one can penalize the model with higher
number of parameters using, e.g. the Akaike information criterion~\cite{Akaike74}, or calculate the Bayes factor by averaging (in the space of parameters) over the full posterior distribution~\cite{Kass95}.

\paragraph*{(3) Validity} 
The probabilistic nature of linguistic laws requires statistical tests. One possible approach
is to compute the probability (p-value) of having observations
similar to the data from a null model compatible with
the linguistic law (which is assumed to be true).  A low p-value is a strong indication that
the null model is violated and may be used to refute the law (e.g., if p-value$<0.01$).
Defining a measure of distance $D$ between the data and the model, the p-value can be
computed as the fraction of finite-size realizations of the model (assuming it is true) that show a
distance $D' > D$. In the case of probability distributions -- linguistic laws of the first kind 
in Sec.~\ref{sec.observations} -- the distance $D$ is usually taken to be the
Kolmogorov-Smirnov distance (the largest distance between the empirical and fitted
cumulative distributions). In the case of simple function -- linguistic laws of the
second and third kind in Sec.~\ref{sec.observations} -- one can consider $D=\chi^2$.

\begin{figure*}
\centering
\includegraphics[width=0.69\columnwidth]{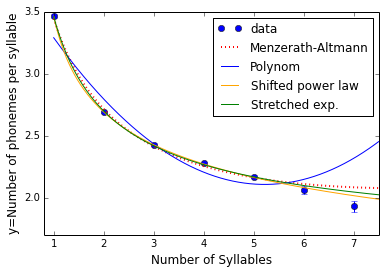}\includegraphics[width=0.715\columnwidth]{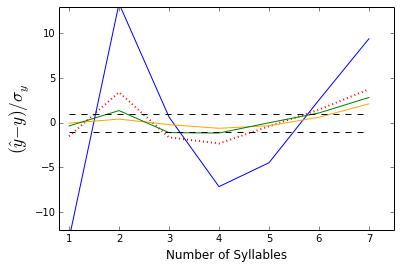}
\caption{Model comparison for the Menzerath-Altmann law. Data points are the
  average over all word (types) in a book  (Moby Dick by H. Melville, as in Fig.~\ref{fig.1}). The curves show the best  fits of the
  four alternative curves, as reported in Tab.~\ref{tab.MA}. Left plot: the data in the
  original scales, as in Fig.~\ref{fig.1}. Right plot: the distance between the curves and
  the points $(\hat{y}-y)/\sigma_y$, where the uncertainty $\sigma_y$ is the standard error of the mean. }
\label{fig.MA}
\end{figure*}

\begin{table*}
  \caption{Likelihood analysis of the Menzerath-Altmann law and three alternative
    functions. The parameters $(\hat{\alpha},\hat{\beta},\hat{\gamma})$ that maximize the
    likelihood $\mathcal{L}_{m}$ of model $m$ were computed using the downhill simplex
    algorithm (using the Python library scipy). The reported p-value corresponds to the fraction of
    random realizations with a $\chi^2$ larger than the observed $\chi^2$. In each realization, 
    one point $y^\dagger(x)$ was generated at each $x$ from a Gaussian distribution
    centered at the model prediction $y_m(x)$ with a standard deviation $\sigma_y(x)$
    given by the data. The best models and the results with $p>0.01$ are shown in bold face.}\label{tab.MA} 
  \centering
  \begin{tabular}{|l|c|c  c c|}
\hline
    {}
&{Menzerath-Altmann (MA)}
&{Shifted power law}
&{Stretched exp.}
&{Polynom}\\
    {}
&{$\alpha x^\beta \exp{(-\gamma x)}$}
&{$\alpha(x+\beta)^\gamma$}
&{$\alpha \exp{(\beta x)}^\gamma$}
&{$\alpha +\beta x+\gamma x^2$}\\
\hline
\multicolumn{5}{|l|}{Results for one book (Moby Dick by H. Melville)} \\
\hline
    {($\hat{\alpha},\hat{\beta},\hat{\gamma}$)}
&{$(3.3,-0.12,-0.051)$}
&{$(2.8,-0.65,-0.19)$}
&{$(1.5,1.4,-0.51)$}
&{$(3.9,-0.69,0.066)$}\\
    {$\log_e \mathcal{L}_{m}/\mathcal{L}_{MA}$}
&{0}
&{{\bf 33}}
&{25}
&{-475}\\
    {p-value}
&{$<10^{-5}$}
&{${\bf 0.611}$}
&{${\bf 0.064}$}
&{$<10^{-5}$}\\
\hline
\multicolumn{5}{|l|}{ Results for English Wikipedia}\\
\hline
    {($\hat{\alpha},\hat{\beta},\hat{\gamma}$)}
&{$(3.2,-0.45,-0.064)$}
&{$(2.8,-0.70,-0.18)$}
&{$(1.6,1.5,-0.60)$}
&{$(3.8,-0.64,0.061)$}\\
    {$\log_{e} \mathcal{L}_{i}/\mathcal{L}_{MA}$}
&{0}
&{$11$}
&{${\bf 49}$}
&{$-1898$}\\
    {p-value}
&{$<10^{-5}$}
&{$ 2\times 	10^{-5}$}
&{${\bf 0.93}$}
&{$<10^{-5}$}\\
\hline
  \end{tabular}
\end{table*}

\paragraph*{Application: Menzerath-Altmann law.} We applied the likelihood analysis summarized above to the case of the Menzerath-Altmann
law introduced in Sec.~\ref{sec.observations}. Our
critical assumption here is that the law is intended to describe the average 
number of phonemes per syllable, $y$, computed over many words $w$ with the same number of
syllables $x$. Assuming the words are independent of each other, the uncertainty in $y(x)$ is thus the standard error of the mean given by
$\sigma_y(x)=\sigma_w(x)/\sqrt{N(x)}$, where $\sigma_w(x)$ is the (empirical) standard
deviation over the words with $x$-syllables and $N(x)$ is the number of such words.

In Fig.~\ref{fig.MA} and Tab.~\ref{tab.MA} we report the fitting, model comparison, and
validity analysis for the Menzerath-Altmann law -- Eq.~(\ref{eq.menzerath}) -- and 
three alternative functions with the same number of parameters. The results show that two
of the three alternative functions (shifted power law and stretched exponential) provide a
better description than the proposed law, which we can safely consider to be incompatible
with the data (p-value$<10^{-5}$). Considering the two databases, the stretched exponential
distribution provides the best description and is not refuted. These
results depend strongly on the procedure used to identify phonemes and syllables (see Appendix).

\subsection{Critical discussion}
In the next paragraphs we critically discuss the likelihood approach considering the example of Zipf's law.

\paragraph*{Fitting as model comparison.} In the beginning of this section we started with the distinction between fitting (i.e., fixing
free parameters) and model comparison (i.e. choosing between different models). This division is
didactic~\cite{Clauset2009}, but from a formal
point of view both procedures correspond to hypothesis testing  because the free
parameters of one fitting model can be thought as
a {\it continuous} parameterization of different models which should be compared and
selected according to their likelihood~\cite{Jaynes}. This means that the points mentioned
below apply equally well to both fitting and hypothesis testing (and, in most cases, also
to test the validity of the models).

\paragraph*{Fitting ranks.}
Power-law fitting recipes~\cite{Clauset2009}--
 employed for linguistic~\cite{Jaeger} and non-linguistic problems --  suggest to fit Zipf's law using
 the distribution of  frequencies $P(f)$ given in Eq.~(\ref{eq.zipf3}).  However, it is
 also possible to use the rank formulation~(\ref{eq.zipf2})~\cite{PRX} because the
 frequency of ranks $f(r)$  is normalized $\sum_r f(r)=1$ and can thus be interpreted as a probability
 distribution.  
However, a drawback in fitting $f(r)$ is that the process of ranking introduces a bias in
the estimator \cite{Gunther,Pietronero}.
For instance, consider a finite sample from a true Zipf distribution containing ranks $r=1, \ldots
,\infty$. Because of statistical fluctuations, some of the rankings will be inverted (or
absent) so that when we rank the words according to the observations obtain ranks
different from the ones drawn. This effect introduces bias in our 
estimation of the parameters (overestimating the quality of the fit). The words affected by this bias are the
ones with largest ranks, which contribute very little to the estimation of the parameters
of Zipf's law (as discussed below). 
Therefore, we expect that this bias to become negligible for sufficiently large sample sizes.

\begin{figure*}[!bt]
\centering
\includegraphics[width=1.6\columnwidth]{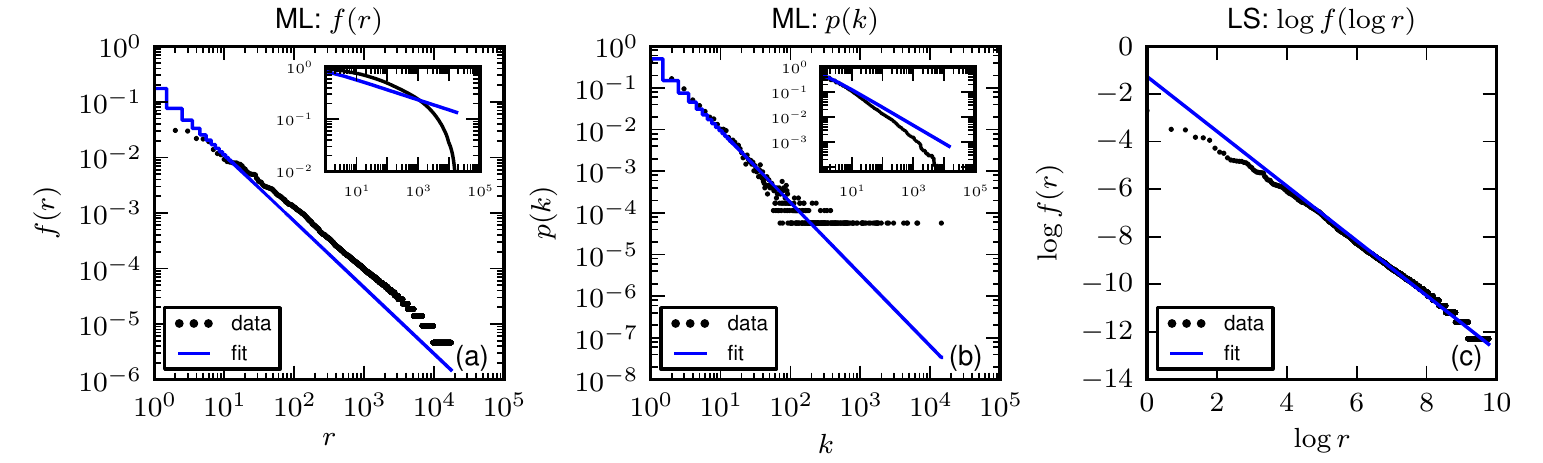}\\
\includegraphics[width=1.6\columnwidth]{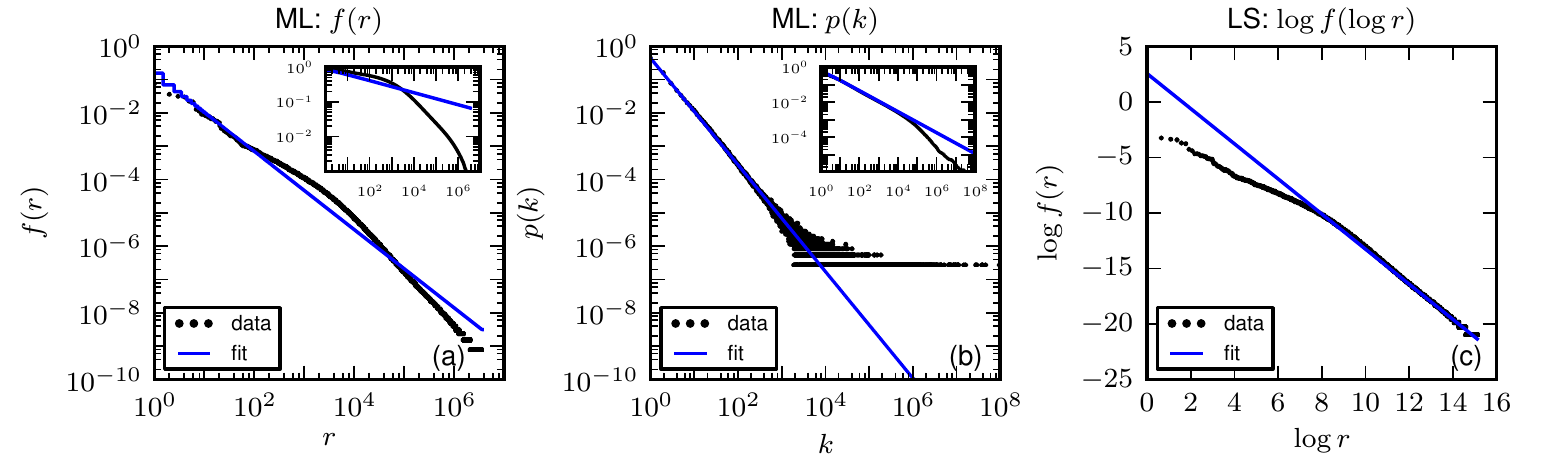}
\caption{Comparison of the Zipf's law obtained using three different fitting
  methods. Results are shown for one book (Moby Dick by H. Melville, top row) and for
  the complete English Wikipedia (bottom row). Data is fitted using Maximum Likelihood
  (ML) in the frequency rank $f(r)$ (left), ML in the frequency distribution $P(f) \sim
  p(k)$ (center), and   least square (LS) in the $\log f$ vs. $\log r$ representation
  (right). 
Insets show the cumulative distributions.
See Tab.~\ref{tab.books}
  for the parameter $\hat{\alpha}$ and significance test of the fits. In the plot in the
  center, instead of $P(f)$ we use  the distribution the unnormalized frequency $p(k)$
  (i.e., $k$ is the number of occurrences of a word in the database).
For ML fits, we used a discrete power law in $f(r)$ and $p(k)$ with support in
$[1,\infty)$ (exponents were obtained using the downhill simplex algorithm of the Python library scipy).  
For the LS fit, we used a continuous straight line in $\log f(\log r)$~\cite{Hastie09}.
}
\label{fig.zipf}
\end{figure*}

\begin{table*}
  \caption{Zipf's law exponent obtained using different fitting methods, see
    Fig.~\ref{fig.zipf}. 
In the fit of $P(f)$ (frequency) we obtain $\hat{\alpha}^\dagger_Z$ and calculate
$\hat{\alpha}_Z=1/(1-\hat{\alpha}^\dagger_Z)$, see Eqs.~(\ref{eq.zipf2})-(\ref{eq.zipf3}).
English version of the books were obtained from the Project Gutenberg, see Appendix. }\label{tab.books}
  \centering
  \begin{tabular}{|l|c  | c|c |  c|c| c|}
\hline
{} & \multicolumn{2}{c|}{Rank: $f(r)$} & \multicolumn{2}{c|}{Frequency: $P(f)$} &
\multicolumn{2}{c|}{Linear: $\log f(\log r)$}\\
{Book 	}&{ $\hat{\alpha}_Z$}&{p-value}&{$\hat{\alpha}_Z$}&{ p-value}&{ $\hat{\alpha}_Z$}&{$R^2$}\\
\hline
{Alice's Adventures in Wonderland (L. Carroll)}&{ 1.22}&{ $<10^{-4}$	}&{ 1.46}&{ $<10^{-4}$	}&{ 1.21}&{ 0.97}\\
{The Voyage Of The Beagle (C. Darwin)}&{ 1.20}&{ $<10^{-4}$	}&{ 1.59}&{ $<10^{-4}$	}&{ 1.29}&{ 0.97}\\
{The Jungle (U. Sinclair)}&{ 1.21}&{ $<10^{-4}$	}&{ 1.45}&{ $<10^{-4}$	}&{ 1.22}&{ 0.98}\\
{Life On The Mississippi (M. Twain)}&{ 1.20}&{ $<10^{-4}$	}&{ 1.38}&{ $<10^{-4}$	}&{ 1.16}&{ 0.98}\\
{Moby Dick; or The Whale (H. Melville)}&{ 1.19}&{ $<10^{-4}$	}&{ 1.38}&{ $<10^{-4}$	}&{ 1.15}&{ 0.98}\\
{Pride and Prejudice (J. Austen)}&{ 1.21}&{ $<10^{-4}$	}&{ 1.66}&{ $<10^{-4}$	}&{ 1.35}&{ 0.98}\\
{Don Quixote (M. Cervantes)}&{ 1.21}&{ $<10^{-4}$	}&{ 1.70}&{ $<10^{-4}$	}&{ 1.38}&{ 0.98}\\
{The Adventures of Tom Sawyer (M. Twain)}&{ 1.21}&{ $<10^{-4}$	}&{ 1.29}&{ $<10^{-4}$	}&{ 1.12}&{ 0.98}\\
{Ulysses (J. Joyce)}&{ 1.18}&{ $<10^{-4}$	}&{ 1.15}&{ $<10^{-4}$	}&{ 1.03}&{ 0.97}\\
{War and Peace (L. Tolstoy)}&{ 1.20}&{ $<10^{-4}$	}&{ 1.84}&{ $<10^{-4}$	}&{ 1.44}&{ 0.97}\\
\hline
{English Wikipedia	}&{ 1.17}&{ $<10^{-4}$	}&{ 1.60}&{ $<10^{-4}$	}&{ 1.58}&{ 0.99}\\
\hline
\end{tabular}
\end{table*}

\paragraph*{Representation matters.}  Equivalent formulations of the linguistic laws lead
to {\it different} statistical analysis and conclusions~\cite{Gunther,Pietronero}.
One example of this point is the use of transformations before the fitting is performed,
such as the linear fit of Zipf's law in logarithmic scale discussed in
Sec.~\ref{ssec.graphical}. The variables used to represent the linguistic law are also
crucial when likelihood methods are used, as discussed above for the case of Zipf's law
represented in $f(r)$ or $P(f)$. 
While asymptotically these formulations are equivalent, the likelihood computed in both
cases is different.
In the likelihood of $P(f)$, an observation corresponds to the frequency of a word {\it
  type}. This means that the most frequent words in the database  count the
same as words appearing only once (the hapax-legomenan).  In practice, the part of the
distribution that matters the most in the fitting (and in the likelihood) are the words with very few counts, which contribute very little to the total text.
In the likelihood of $f(r)$ the observational quantity is the rank $r$ of each occurrence of
the word meaning that each word {\it token} counts the same. This means that the frequent
words contribute more and the fitting of $f(r)$ is robust against rare words.
Linear regression in log-log plot counts every point in the plot the same and, since  there are more
points for large $r$,  low-frequency words dominate the fit.  Using logarithmic binning,
as suggested in Ref.~\cite{Goldstein04}, equalize the importance of words across $log(r)$.
In summary, while fitting a straight line in log-log scale using logarithmic binning gives
the same value for words across the full 
spectrum (in a logarithmic scale), the statistical rigorous methods of Maximum Likelihood will be dominated either by the most frequent (in case of fitting in $f(r)$) or least frequent (in case of fitting in $P(f)$) words.

Beyond Zipf's law,  the reasoning above shows that even if asymptotically (i.e. infinite
data) different formulations of a law are equivalent, the  representation in which we test
the law matters because it assumes a sampling process of the data.
This in turn leads to different results when applied to finite and often noisy data and
has to be taken into account when interpreting the results.

\paragraph*{Application: Fitting Zipf's law.}
In Fig.~\ref{fig.zipf} and Tab.~\ref{tab.books} we compare the different fitting methods
described above.  
The visual agreement between data and the fitted curves reflects the different
weights given by the methods to different regions of the distribution as discussed above (high-frequency
words for $f(r)$ and low-frequency words for the other two cases).  Not surprisingly,
Tab.~\ref{tab.books} shows that the estimated exponent $\alpha$ varies from method to
method. This variation is larger than the variation across different databases.  Large values of $R^2$
computed in the linear fit, usually interpreted as an indication of good fitting, are observed also when
the p-value are very low.

\begin{figure*}[!bt]
\centering
\includegraphics[width=0.8\columnwidth]{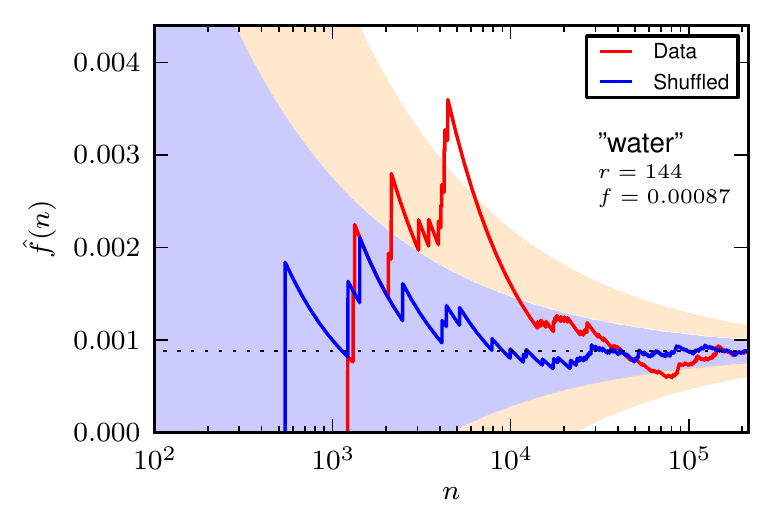}\includegraphics[width=0.8\columnwidth]{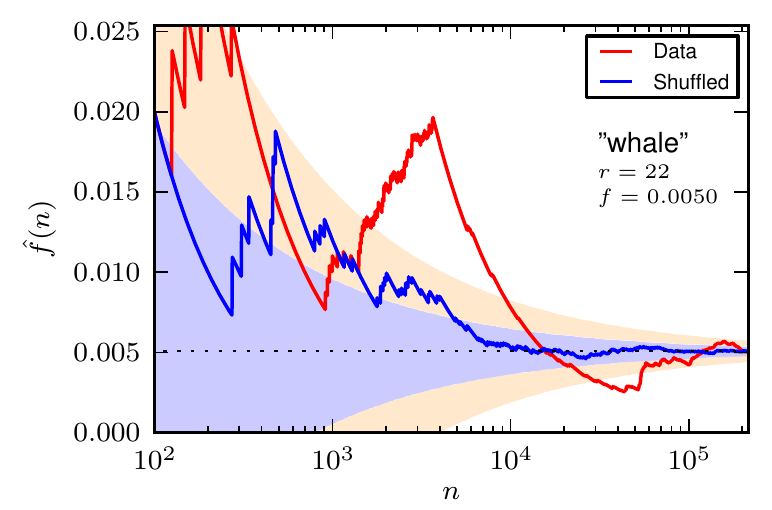}
\caption{Estimation of the frequency of a word in the first $n$ word tokens of a book
  (Moby Dick by H. Melville). The red curve corresponds to the actual observation 
(word ``water'' in the left and word ``whale'' in the right)
and the blue curve to
  the curve measured in a version of the book in which all word tokens were randomly
  shuffled. The shaded regions show the expected fluctuations ($\pm 2 \sigma$) assuming that
  the probability of using the word is given by the frequency of the word in the whole
  book ($f(n=N)$ and that: (i) usage is random (blue region) -- see also Ref.~\cite{Baayen}  or (ii) the time between
  successive usages of the word is drawn randomly from a stretched exponential
  distribution with exponent $\beta=0.5$, as proposed in Ref.~\cite{Altmann2009}.}
\label{fig.4}
\end{figure*}

\paragraph*{Correlated samples}  The failure of passing significance tests for increasing
data size is not surprising because any small deviation from the null model becomes statistically significant. 
A possible conclusion emerging from these analysis is that power-law distributions
are not as widely valid as previously claimed (see also Refs.~\cite{CriticalTruth,Clauset2009}), but often are better 
than alternative (simple) descriptions (see our previous publication Ref.~\cite{PRX}  in
which we consider two-parameter generalizations of Zipf's law).
The main criticism we have on this widely used framework of analysis is that it
ignores the presence of correlations in the data: the computation of the likelihood in Eq.~(\ref{eq.likelihood})
assumes independent observations. Furthermore, this assumption leads to an underestimation of the expected fluctuations (e.g. KS-distance) in the calculation of the p-value when assessing the validity of the law. 
It is thus unclear in which extent a negative result in the validity test (e.g.,
p-value$\ll 0.01$) is due to a failure of the proposed law or, instead, is due to the violation of the
hypothesis of {\it independent} sampling. This hypothesis is known to be violated in
texts~\cite{Baayen,Lijffijt2011}: the 
sequence of words and letters are obviously related to each other. In Fig.~\ref{fig.4} we
show that these correlations affect the estimation of the frequency of individual words,
which show fluctuations much larger than those expected not only based on the independent
random usage of words (Poisson or bag of word models) but also in a null model in which
burstiness is included~\cite{Altmann2009,Lijffijt2011}. Altogether, this shows that the independence assumption -- used to
write the likelihood~(\ref{eq.likelihood}) -- is strongly violated and affects both
the analysis based on $f(r)$ (correlation throughout a book) and $P(f)$ ($f$ can be
thought as a finite size estimation as the ones shown in Fig.~\ref{fig.4}).

One approach to take into account correlations is to estimate a time for which two observations are
independent, and then consider observations only after this time (a smaller effective
sample size). Alternative approaches considered statistical tests for specific classes of
stochastic processes (correlated in time)~\cite{Weiss78} or based on estimations of the
correlation coming from the data~\cite{Chicheportiche11}. The application of these methods
to linguistic laws is not straightforward because these methods fail in
cases in which no characteristic correlation time exist. Books show such long-range
correlations~\cite{Schenkel93},  also in the position of individual words in
books~\cite{Mandelbrot73,PNAS}, in agreement with the observations reported in
Fig.~\ref{fig.4}. More generally, correlations lead to a slower convergence to asymptotic
values and it is thus  possible 
to create processes of text generation that comply to a linguistic law asymptotically but
that (in finite samples) violate statistical tests based on independent sampling. The
problem affects also model comparison and fitting because these problems are also based on
the likelihood (in these cases, correlation affects all models and therefore
it is unclear the extent in which it impacts the choice of the best model).

\section{Relation between laws}\label{sec.relation}

\begin{figure*}[!bt]
\centering
\includegraphics[width=1.6\columnwidth]{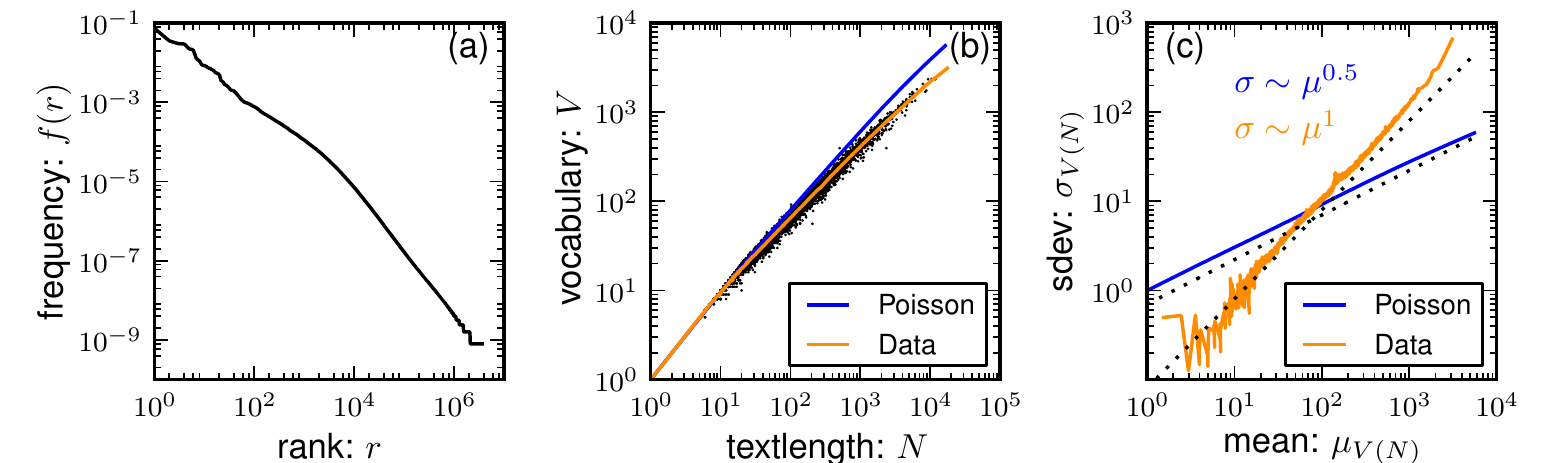}
\caption{Relation between Zipf's law and Heaps' law in the English Wikipedia. Fixing the
  rank-frequency distribution of the complete English Wikipedia -- shown in panel (a) --
  and assuming each word to follow a Poisson process (i.e., to be used randomly) with
  fixed frequency $f(r)$, one obtains the blue curve for 
  the Heaps' law in (b). Considering each Wikipedia article separately -- as shown by
  black dots  in (b) -- we estimate in a
  moving window centered in $N$ the average $\mu_{V(N)}$  and standard deviation
  $\sigma_{V(N)}$ over all articles in the window. The dependence of $\mu_V(N)$ on $N$ is
  shown in (b) by a solid line. The dependence of $\sigma_V(N)$ on $\mu_V(N)$ is shown in
  (c) and reveals a different scaling than the one predicted by the Poisson model. Figure
  adapted from Ref.~\cite{NJP}.}
\label{fig.5}
\end{figure*}

In view of the different laws proposed to describe text properties, a natural question is
the relationship between them (e.g., whether one law can be derived from another or
whether there are generative processes that account for more than one law
simultaneously). For instance, Ref.~\cite{PNAS} clarifies 
how the long-range correlation of texts is related to the skewed distribution of
recurrence time between words~\cite{Zipf,Altmann2009,Lijffijt2011} (a consequence of burstiness~\cite{Zanette,Serrano2009}). Another well-known
relation is the connection between Heaps' law and Zipf's law~\cite{Lu,Bernhardsson2010,PRX,Francesc,NJP} (see
Refs. \cite{Ebeling1994,Debowski2006,Serrano2009,NJP} for other examples). Here again
the importance of fluctuations and an underlying null model is often neglected.

The need for a null model is evident if we consider a text in which all possible words
appear once in the very beginning of the text, violating  Heaps' law, even though their frequency over the full text is still
compatible with Zipf's law. A typical null model is to consider that every word is used
independently from the others with a probability equal to its global frequency. This
probability is usually taken to be constant throughout the text (Poisson process), but alternative
formulations considering time-dependent frequencies lead to similar results. For this 
generative model, Zipf's law~(\ref{eq.zipf2}) leads to a Heaps' law~(\ref{eq.heaps}) with parameters $\alpha_H=
1/\alpha_Z$~\cite{PRX}. Similar null models are implicitly or explicitly assumed
in different derivations~\cite{Lu,Bernhardsson2010,PRX,Francesc,NJP}.

Figure~\ref{fig.5} shows that the connection between Zipf's and Heaps' law using the
independent usage of words fails to reproduce the fluctuations observed in data. 
In particular, the fluctuations around the average vocabulary size $V$ predicted from Heaps'
law scales linearly with $N$, and not as $\sqrt{N}$ as predicted by the independence assumption
(through the central limit theorem). In Ref.~\cite{NJP} we have shown that this scaling
-- also known as Taylor's law~\cite{Eisler08} -- is a result of correlations in the usage among different words
induced by the existence of topical structures inside and across books.

\section{Discussion}

It is common to find claims that a particular linguistic law is valid in a language or
corpus. A closer inspection for the statistical support of these claims is often
disappointing.  In this Chapter we performed a critical discussion of linguistic laws, the
sense in which they can be considered valid, and the extent in which the evidence support
its validity. We argued that linguistic laws have to be interpreted in a statistical
sense. Therefore,  model selection (also fitting) and the compatibility to data have to be
performed computing statistical tests based on the likelihood (plausibility) of the
observations. 
The statistical analysis is far from being free of choices, both in terms of the methods
employed and also about additional assumptions not contained in the original law, as
discussed below. The analysis we presented above is intended to show that 
these choices matter and should be carefully discussed.
The picture that emerges from the straight applications of the statistical tests above is
that: (i) the linguistic laws are often the best simple description of the data, but (ii) the data 
is not generated according to it so that in a strict sense the validity of the law is
falsified. This interpretation suggests that linguistic laws are useful and capture some
of the ingredients seen in language, but are unable to describe observations in full
detail even in the limit of large texts (possibly because of the existence of additional
processes ignored by the law). 

The main limitation of the methods we described, and thus of the conclusions summarized
above, is that they were based not only on the
statement of the law but also on the hypothesis that
observations are independent and identically distributed. This hypothesis is known to be
violated in almost all observations of written language. It is thus
unclear in which extent the rejection of the null model (small p-value) can be considered
a falsification of the linguistic law. On the one hand, this reasoning shows the
limitation of the 
statistical methods and the necessity to apply and develop tests able to deal with (long-range)
correlated data. On the other hand, it shows that the usual statements of linguistic laws
are incomplete because they 
cannot be properly tested.  A meaningful formulation of a linguistic law allows for the
computation of the likelihood of the observations, e.g., it should be 
accompanied by a prediction of the fluctuations, a generative model for the relevant
variables,  or, ultimately, a model for the generation of texts.
Such models are usually interpreted as an explanation of the origin of the
laws~\cite{Mitzenmacher2004,Newman,Piantadosi2014} and are absent from the
statement of the linguistic laws, despite the 
fact that Herdan already drew attention to this point~\cite{Herdan}: {\it
``The quantities which we call statistical laws being only expectations, they are subject
to random fluctuations whose extent must be regarded as part of the statistical law.''}
In the same sense that a scientific law cannot be
judged separated from a theory,  linguistic laws are only fully defined once a generative
process is given. 
The existence of long-range correlations, burstiness, and
topical variations lead to strong fluctuations in the estimations of observables in texts,
including the quantities described by linguistic laws. 

Our findings have consequences to applications in information retrieval and text
generation. For instance, our results show that strong fluctuations around specific
laws are observed and that results obtained using the independence assumption (e.g.,
bag-of-words models) have a limited applicability. Therefore, statistical laws should not be imposed too strictly in the
generation of artificial texts or in the analysis of unknown databases. Large fluctuations
are as much a characteristic of language as the laws themselves and therefore the
creativity in the generation of texts is much larger than the one obtained if laws are
imposed as strict constraints.

Finally we would like to mention that our conclusions apply also to other statistical
laws beyond linguistic. Invariably, the increase of data size leads to a rejection of 
null-models, e.g.  many recent works emphasize that claims of power-law distributions  do
not survive rigorous statistical tests~\cite{Li,Clauset2009,CriticalTruth}. However, the
statistical tests employed in these references, 
and in most likelihood-based analysis, rely on the independence assumption of the
observations (known to be violated in many of the treated cases).
Nevertheless, we are not aware that this point has been critically discussed in the large
number of publications on power-law fitting. The crucial role of mechanistic models in
the fitting and statistical analysis of scaling laws was emphasized in Ref.~\cite{smog}
for urban-economic data. 

{\bf Acknowledgments:} We thank Roger Guimer\`a, Francesc Font-Clos, and Anna Deluca for insightful discussions.

\section*{Appendix}

The books listed in Tab.~\ref{tab.books} were obtained from Project
Gutenberg (\url{http://www.gutenberg.org}). The books and data filtering are the same as
the ones used in Ref.~\cite{PNAS} (see the Supplementary information of that paper for
further details).  We removed capitalization and all symbols except the letters ``a-z'', the number ``0-9'',
the apostrophe, and the blank space. A string of symbols between two consecutive blank
spaces was considered to be a word. 

The English Wikipedia data was obtained from Wikimedia dumps (\url{http://dumps.wikimedia.org/}).
The filtering was the same as the one used in Ref.~\cite{NJP}, in which we removed capitalization and kept only those words (i.e. sequences of symbols separated by blank space) which consisted exclusively of the letters ``a-z'' and the apostrophe.

The computation of Menzerath-Altmann law appearing in
Figs.~\ref{fig.1},~\ref{fig.MA}, and Tab.~\ref{tab.MA} was done starting from the unique
words (word type) in the database discussed in the previous paragraphs. For each word $w$ we
applied the following steps: 

\begin{enumerate}
\item Lemmatize using the WordNetLemmatizer (\url{http://wordnet.princeton.edu}
  in the NLTK Python package \url{http://www.nltk.org/}) . 
\item Count the number of syllables $x_w$ based on the {\it Moby Hyphenation List by Grady
    Ward}, available at \url{http://www.gutenberg.org/ebooks/3204}
\item Count the number of phonemes $z_w$ based on {\it The CMU Pronouncing Dictionary},
  version 0.7b available at \url{www.speech.cs.cmu.edu/cgi-bin/cmudict}
\end{enumerate}
For the book {\it Moby Dick} by H. Melville, this procedure allowed to compute $x_w$ and $z_w$ for $11,595$
words, $66\%$ of the total number of words (before lemmatization). For the Wikipedia, we
obtain $60,749$ words,  $1.7\%$ of the total number. The low success in Wikipedia is due
to the size of the database (large number of rare words) and the results depend more
strongly on the procedure described above than on the database itself.

%
%

\begin{thebibliography}{99.}%
%
%

\bibitem{Herdan} G. Herdan, Quantitative Linguistics (Butterworth Press, Oxford, 1964)


\bibitem{Zipf} G. K.  Zipf, The Psycho-Biology of Language  (Routledge,
  London, 1936). {\it Id.},  Human behavior and the principle of least effort (
  Addison-Wesley Press, Oxford, 1949).   


\bibitem{QL} R. K\"ohler, G.  Altmann, and R. G.  Piotrowski (Eds.)
  Quantitative Linguistik. Ein internationales Handbuch. Quantitative Linguistics. An
  international Handbook. (=HSK27) (de Gruyter, Berlin, 2005). 

\bibitem{Series} R. K\"ohler, G.  Altmann, and P. Grzybek (Eds), Quantitative Linguistics,
  De Gruyer Mouton,  \url{www.degruyter.com/view/serial/35295} Cited 6 Feb 2015.

\bibitem{Glottopedia} Glottopedia: the free Encyclopedia of Linguistics,   \url{http://www.glottopedia.org/index.php/Laws}  Cited 17 Dec 2014.

\bibitem{TrierWebpage} Enciclopedia entry: {\it Laws in Quantitative linguistics},
  \url{http://lql.uni-trier.de} Cited 3 Dec 2014

\bibitem{Baayen} R. Harald Baayen, Word Frequency Distributions, Kluwer Academic Pub.,
  Dordrecht (2001).

\bibitem{Zanette} D. H. Zanette, Statistical Patterns in Written Language, arXiv:1412.3336 (2014).


\bibitem{Barbieri} G. Barbieri, F. Pachet, P. Roy, and M. Degli Esposti, Markov
  Constraints for Generating Lyrics with Style, 20th European Conference on Artificial
  Inteligence -- ECAI (IOS Press, Amsterdam, 2012)


\bibitem{Simon} H. A. Simon, On a Class of Skew Distribution Functions, Biometrika 42, 425 (1955).

\bibitem{Mitzenmacher2004} M. Mitzenmacher, A Brief History of Generative Models for Power
  Law and Lognormal Distributions, Internet Mathematics 1, 226 (2004).

\bibitem{Newman} M. E. J. Newman, Power Laws, Pareto Distributions and Zipf’s law,
  Contemp. Phys. 46, 323 (2005).


\bibitem{Li} W. Li, Zipf's Law Everywhere, Glottometrics 5, 14 (2002).

\bibitem{Piantadosi2014} S.T. Piantadosi, Zipf’s word frequency law in natural language: A
  critical review and future directions. Psychonomic bulletin \& review, 21,1112  (2014).

\bibitem{Montemurro} D. Zanette and M. Montemurro, Dynamics of Text Generation with
  Realistic Zipf's Distribution, J. Quant.  Linguist. 12, 29 (2005).

\bibitem{Altmann1980} G. Altmann, Prolegomena to Menzerath's law, Glottometrika 2,
  1 (1980).

\bibitem{Cramer2005}  I. Cramer. The Parameters of the Altmann-Menzerath Law. J.
  Quant. Linguist. 12, 41 (2005).





\bibitem{Egghe} L. Egghe, Untangling Herdan's Law and Heaps' Law : Mathematical and
  Informetric Arguments, J. Am. Soc. Inf. Sci. Tec. 58,702 (2007).


\bibitem{Lu} L. L\"u, Z.-K. Zhang, T. Zhou, Zipf's Law leads to Heaps' law: Analyzing
  Their Relation in Finite-Size Systems, PLOS ONE 5, e14139 (2010).

\bibitem{Petersen} A. M. Petersen, J. N. Tenenbaum, S. Havlin, H. E. Stanley, and M. Perc,
  Languages cool as they expand: Allometric scaling and the decreasing need for new
  words, Scientific Reports 2, 943 (2012)

\bibitem{PRX} M. Gerlach and E. G. Altmann, Stochastic model for the vocabulary growth in
  natural languages, Physical Review X 3, 021006 (2013)

\bibitem{Francesc} F. Font-Clos, G. Boleda, A. Corral, A scaling law beyond Zipf's law and
  its relation to Heaps' law,  New Journal of Physics, 2013.

\bibitem{NJP} M Gerlach and EG Altmann, Scaling laws and fluctuations in the statistics of word frequencies, New Journal of Physics 16, 113010 (2014)

\bibitem{Altmann2009} E. G. Altmann, J. B. Pierrehumbert, and A. E. Motter. Beyond word frequency: bursts, lulls, and scaling in the temporal distributions of words. PlosOne. 4, e7678 (2009).


\bibitem{Lijffijt2011} J. Lijffijt,  P. Papapetrou, K. Puolam\"aki, H. Mannila,  Analyzing
  Word Frequencies in Large Text Corpora Using Inter-arrival Times and Bootstrapping,
  Chapter in Machine Learning and Knowledge Discovery in Databases, Lecture Notes in
  Computer Science 6912, 341 (2011).






\bibitem{Mandelbrot73}
 F.J. Damerau, B. Mandelbrot, Tests of the degree of word clustering in samples of written
 English, Linguistics 102, pp. 58-72 (1973).

\bibitem{Schenkel93} A. Schenkel, J. Zhang, and Y. Zhang, Long range correlation in human
  writings, Fract. 1, 47 (1993).


\bibitem{PNAS}  E. G. Altmann, G. Cristadoro, and M. Degli Esposti. On the origin of
  long-range correlations in texts. PNAS. 109, 11582  (2012).



\bibitem{Ebeling1994} W. Ebeling and T. Pöschel Europhys. Lett. 26 24 (1994).

\bibitem{Debowski2006} 
L. Debowski. On Hilberg's law and its links with Guiraud's law. J. Quant. Linguist.. 13, 81-109 (2006).


  

\bibitem{Piantadosi2011}    S. T. Piantadosi,  H. Tily, and E. Gibson, Word lengths are optimized for efficient
    communication, PNAS 108, 3526 (2011).

\bibitem{Sole10} R. V. Sol\'e, B. Corominas-Murtra, S. Valverde, and  L. Steels, Language
  Networks: their structure, function and evolution, Complexity 15, 20 (2009)

\bibitem{Choudhurry09} M. Choudhury and A. Mukherjee: The Structure and Dynamics of
  Linguistic Networks,  in Dynamics On and Of Complex
  Networks (Springer 2009) pp 145-166.

\bibitem{Baronchelli13} A. Baronchelli, R. Ferrer-i-Cancho, R. Pastor-Satorras, N. Chater,
  M. H. Christiansen, Networks in Cognitive Science, Trends in Cognitive Sciences 17, 348
  (2013). 

\bibitem{Cong14} J. Cong and H. Liu, Approaching human language with complex networks,
  Physics of Life Reviews 11, 598 (2014).





\bibitem{Mandelbrot61} 
B. Mandelbrot, 1961 {On the Theory of Word Frequencies and on Related Markovian Models of Discourse} {\em
  Structure of Language and Its Mathematical Aspects: Proceedings of Symposia
  in Applied Mathematics Vol. XII\/} (Providence, RI: American
  Mathematical Society)


\bibitem{WikipediaConstrainedWriting} Constrained Writing, in \textit{Wikipedia}, \url{http://en.wikipedia.org/wiki/Constrained_writing}   Cited 3 Dec 2014.

\bibitem{Amancio} D. R. Amancio, E. G. Altmann, D. Rybski, O. N. Oliveira Jr, LdF
  Costa, Probing the Statistical Properties of Unknown Texts: Application to the Voynich
  Manuscript,   PLOS ONE 8, e67310 (2013).


\bibitem{Bernhardsson2010} S. Bernhardsson, L. E. C. da Rocha,  P. Minnhagen,  Size-dependent
  word frequencies and translational invariance of books, Physica A 389, 330 (2010).

\bibitem{Ger} G. Febres, K. Jaff\'e, and C. Gershenson, Complexity Measurement of Natural
  and Artificial Languages, Complexity doi:10.1002/cplx.21529 (2014)

\bibitem{Dodds} J.R. Williams, J.P. Bagrow, C.M. Danforth, and P.S. Dodds, Text mixing shapes the
  anatomy of rank-frequency distributions: a modern zipfian mechanics for natural
  language, arXiv:1409.3870 (2014)

\bibitem{Baixeries}  J. Baixeries,     B. Elvevag,     R. Ferrer-i-Cancho, The
  Evolution of the Exponent of Zipf's Law in Language Ontogeny, PLOS ONE 8, e53227 (2013).

\bibitem{Jaeger} G. J\"ager, Power Laws and other heavy-tailed distribution in linguistic
  typology, Adv. Comp. Syst. 15, 1150019 (2012).

\bibitem{Corominas-Murtra} B. Corominas-Murtra, J. Fortuny, and R. V. Solé, Emergence of
  Zipf’s Law in the Evolution of Communication, Phys. Rev. E 83, 036115 (2011).

\bibitem{Ferrer-i-Cancho14} R. Ferrer-i-Cancho, Optimization models of natural
  communication, arXiv:1412.2486 (2014).



\bibitem{Marsili13} M. Marsili, I. Mastromatteo, and Y. Roudi,  On sampling and modeling
  complex systems, J. of Stat. Mech.  2013,  P09003 (2013).

\bibitem{Peterson13} J. Peterson, P. D. Dixit, and K. Dill, A maximum entropy framework for
nonexponential distributions, PNAS, 110, 20380 (2013)

%

\bibitem{Goldstein04}
M. L. Goldstein, S. A. Morris, and G. G. Yen. Problems with fitting to the power-law distribution. European Journal of Physics B. 41, 255-258 (2004). 

\bibitem{Bauke07}
H. Bauke. Parameter estimation for power-law distributions by maximum likelihood methods. European Journal of Physics B. 58, 167-173 (2007). 

\bibitem{Clauset2009} A. Clauset, C. R.Shalizi, and M. E. J. Newman. Power-Law Distributions in Empirical Data. SIAM Review. 51, 661–703 (2009).


\bibitem{Deluca13}
A. Deluca and A. Corral. Fitting and goodness-of-fit test of non-truncated and truncated power-law distributions. Acta Geophysica. 61, 1351-1394 (2013).

\bibitem{Hastie09}
T. Hastie, R. Tibshirani, and J. Friedman. The elements of statistical learning (Springer, New York, 2009).


\bibitem{Burnham02}
K. P. Burnham and D. R. Anderson. Model selection and multimodal inference: a practical information-theoretic approach. Spinger, New York (2002).

\bibitem{Akaike74}
H. Akaike. A new look at the statistical model identification. IEEE Transactions on Automatic Control. 19, 716-723 (1974).

\bibitem{Kass95}
R. E. Kass and A. E. Raftery. Bayes Factors. Journal of the American Statistical Association. 90, 773-795 (1995).

\bibitem{Jaynes} E. T. Jaynes, Probability Theory: The Logic of Science (Oxford University
  Press, Oxford, 2003).

\bibitem{Gunther}   R. G\"unther, L. Levitin, B. Schapiro, P. Wagner, Zipf 's law and the
  effect of ranking on probability distributions, Int. J. Theor. Phys. 35, 395 (1996). 

\bibitem{Pietronero} M. Cristelli, M. Batty, and L. Pietronero, There is more than a power
  law in Zipf, Scientific reports  2, 812. (2015). 

\bibitem{CriticalTruth} M.P.H. Stumpf and M. A. Porter. Critical Truths About Power Laws. Science, 335, 665–666 (2012).
  
\bibitem{Weiss78}
M.S. Weiss. Modification of the Kolmogorov-Smirnov Statistic for Use with Correlated Data. Journal of the American Statistical Association. 73, 872-875 (1978).

\bibitem{Chicheportiche11}
R. Chicheportiche and J.-P. Bouchaud. Goodness-of-Fit tests with Dependent Observations. Journal of Statistical Mechanics: Theory and Experiment. P09003 (2011).

\bibitem{Serrano2009} M.A. Serrano, A. Flammini, and F. Menczer, Modeling statistical
  properties of written text., PlOS ONE 4, e5372 (2009).

\bibitem{Eisler08}
Z. Eisler, I. Bartos, and J. Kert\'esz. Fluctuation scaling in complex systems: Taylor's law and beyond. Advances in Physics. 57, 89-142 (2008).

\bibitem{smog}  R. Louf and M. Barthelemy, Scaling: lost in the smog, Environment and
  Planning B: Planning and Design 41, 767 (2014).

\end{thebibliography}
%

\end{document}